\documentclass[11pt,a4paper]{article}
\usepackage[utf8]{inputenc}
\usepackage{amsmath}
\usepackage{placeins}
\usepackage{url}
\usepackage{natbib}
\usepackage{color,xcolor,setspace,geometry}
\setcitestyle{authoryear,open={(},close={)}}
\usepackage{cite}
\usepackage{amsfonts}
\usepackage[normalem]{ulem}
\usepackage{pifont}

\usepackage{arydshln}

\usepackage{tikz}
\usetikzlibrary{shapes}
\usetikzlibrary{plotmarks}
\usetikzlibrary{tikzmark,matrix,calc}
\usetikzlibrary{arrows.meta,calc,decorations.markings,math,arrows.meta}

\usepackage{amssymb}
\usepackage{multirow}
\usepackage{graphicx}
\usepackage{soul}
\soulregister\citep7
\soulregister\cite7
\soulregister\citet7
\soulregister\citealp7
\soulregister\ref7
\usepackage{rotating}
\usepackage{setspace}
\usepackage{threeparttable}
\usepackage{authblk}
\usepackage{subfigure}

\usepackage{appendix}
\usepackage{bm}

\begin{document}
\begin{spacing}{1.5}	

\title{Do Betting Markets Sense a Goal Coming?\\ Evidence from the German Bundesliga}

\author[$\dag$ $\star$]{David Winkelmann}
\author[$\ddag$]{Christian Deutscher}

\affil[$\dag$]{Department of Business Administration and Economics\\ Bielefeld University, Bielefeld, Germany}
\affil[$\ddag$]{Department of Psychology and Sports Science\\ Bielefeld University, Bielefeld, Germany}
\affil[$\star$]{Corresponding author: david.winkelmann@uni-bielefeld.de}

\maketitle
\noindent

\begin{abstract}
We use the fertile ground of betting markets to study the anticipation of major news in financial markets. While there is a considerable body of literature on the accuracy and efficiency of betting markets after important in-match events, there are no studies dealing with the anticipation of such events. This paper tracks bookmaker odds and betting stakes to provide insights into the movement of both prior to goals. Utilising high-resolution (1 Hz) data from a leading European bookmaker for a full season of the top German football league, we analyse whether market participants anticipate major news. In particular, we consider the case of the first goal scored within a match, with its strong impact on the match outcome. Using regression models and state-space models (SSMs) accounting for an underlying market activity level, we investigate whether the bookmaker adjusts odds and bettors tend to place higher stakes on the scoring team right before the first goal is scored. Our results indicate that neither side of the market anticipates goals by significantly adjusting their behaviour.

\end{abstract}

\section{Introduction}
\label{sec:introduction}
Sports betting markets hold significant economic relevance, with a turn-over exceeding 40 billion euros in Europe alone in 2021 \citep{michels2023bettors}. Given the increased competition in recent years and the resulting narrower margins, bookmakers have to be excellent predictors of match outcomes to remain profitable \citep{che2017price, winkelmann2024betting}. Such preciseness is particularly crucial for the growing live-betting market, where bets are placed during an ongoing match. Here, bookmakers must promptly adjust odds in response to in-match dynamics and major news, such as goals \citep{otting2024demand}.

Sports betting markets are similar to general stock markets. Specifically, placing a bet is akin to buying a company's stock \citep{sauer1998economics}. However, in contrast to financial markets, the value of the uncertain asset (the outcome of a bet) becomes known after a fixed deadline \citep{thaler1988anomalies}. Recent financial market literature examines the effect of news on stock prices (see e.g.\ \citealp{chua2019information}, and \citealp{haroon2020covid}). Studies suggest that news sentiment and media coverage can predict stock price movements and volatility. On the one hand, positive news lead to quick increases in stock returns, while on the other hand, negative news causes delayed reactions \citep{Heston2017News}. Incorporating financial news data generally improves prediction accuracy compared to using stock features alone \citep{Dahal2023A}. Market news is perceived differently by investors \citep{Ben-Rephael2017It}, and even fake news can impact stock markets \citep{Clarke2019Fake}. In financial markets, stock price movements before major corporate decisions may result from both insider trading and market anticipation \citep{jarrell1989stock,jain2014stock,tunyi2021revisiting}. In contrast, news in betting markets typically become unambiguously observable to all market participants at the same point in time. Although this provides an excellent environment for analysing the impact of news on market behaviour, to date, there is only limited research in this area. Some indication on the efficiency of betting markets after major events comes from studies using betting exchange data (see e.g.\ \citealp{gil2007testing, choi2014role}, and \citealp{croxson2014information}). Relying on live-betting markets, \citet{otting2024demand} show that bettors tend to overreact to news in football. However, we are unaware of any study investigating the anticipation of in-match news in sports betting markets.

Given the characteristics of sports betting markets as fertile ground for analysing arbitrage opportunities \citep{miller2013intra}, in-match data allows to accurately analyse the anticipation of news by market participants. As football is a low-scoring sport, the first goal is crucial to the match outcome \citep{Lago‐Peñas2016Home}. Therefore, this event is of particular interest when considering major news in football. Studying the anticipation of goals can reveal whether betting markets are rational or contain inefficiencies that skilled bettors can exploit. If odds do not fully reflect the probability of an impending goal, then market participants may have arbitrage opportunities. For example, if bettors systematically anticipate goals before they occur (due to tactical momentum or match dynamics), this would suggest that bookmakers adjust their odds reactively rather than proactively.

Our study relies on in-match data from the German Bundesliga season 2018/19 provided by a large European bookmaker. This unique data features a high resolution of 1 Hz on the bookmaker's odds and stakes placed by bettors. Additionally, it includes information on the precise time of major events during the match, such as goals and red cards. We consider only matches with at least one goal and focus particularly on the market behaviour of the bookmaker and bettors just before the first goal of a match to capture potential anticipation effects. If bettors were to anticipate the first goal, they could generate positive returns by betting on the team that eventually scores the goal. Highly dynamic markets, such as both the financial market and the sports betting market, involve serial correlation in the not directly observable market activity level. This naturally translates to a state-space modelling (SSM) approach, as it was previously suggested by e.g.\ \citet{choi2014role,croxson2014information,otting2024demand} for sports betting markets and e.g.\ \citet{jacquier2002bayesian,al2011identification} for general financial markets. This allows to relate observed relative stakes to a (latent) state of the market.

The remainder of the paper is structured as follows: Section~\ref{sec:data} details our data. In Section~\ref{sec:bookmakers}, we examine the anticipating behaviour of the bookmaker, while Section~\ref{sec:bettors} considers the perspective of bettors. Finally, Section~\ref{sec:conclusion} concludes the paper.

\section{Data}
\label{sec:data}

The data consists of pre-match and in-match betting odds for home wins, draws, and away wins, along with detailed records of stakes placed by bettors for all 306 matches from the 2018/19 Bundesliga season. The 1 Hz resolution of the betting activity results in strong volatility in stakes over time. We thus aggregate the data into one-minute intervals to mitigate such noise in the observed stakes. To focus on the odds movement and betting activity as indicators of anticipating the match's first goal, we exclude all 17 scoreless matches from the sample. Descriptive statistics for the 289 matches featuring at least one goal highlight the importance of the first goal for the final outcome. In 204 matches, the team scoring the first goal ultimately won (70.6\%), while the opposing team won 29 matches (10.0\%), and there were 56 draws (19.4\%).

Football matches consist of two 45-minute halves, a half-time break of about 15 minutes, and typically a few minutes of injury time. The longer a match remains scoreless, the higher the chances for a draw and the lower the chances of a win by any team. Therefore, betting odds crucially depend on the time left to play. We denote the minutes elapsed since the kick-off by $t$, and the minute of the first goal in match $i$ by $T_i$.\footnote{Note that for the first half, $t$ always corresponds to the actual minute of the match. Due to variation in the first half's injury time, the second half starts at approximately minute $t = 60$.} 

\begin{figure}[ht]
    \centering
    \includegraphics[width=0.7\linewidth]{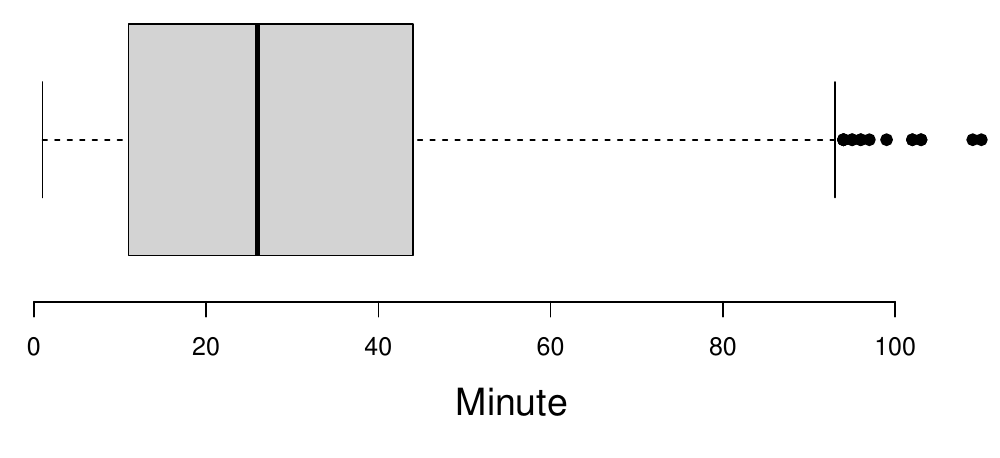}
    \caption{Boxplot of the minute $T_i$ in which the first goal was scored in the 289 non-scoreless matches of the 2018/19 German Bundesliga season.}
    \label{fig:boxplot_minute_firstgoal}
\end{figure}

Figure~\ref{fig:boxplot_minute_firstgoal} shows a boxplot of the minute the opening goal was scored. In our sample, the first goal was scored anywhere between the first minute of a match and late during injury time of the second half. Approximately 75\% of the goals were scored during the first half. The variable $\textit{mintogoal}_{it} = T_i - t$ indicates the time remaining between the current minute $t$ and the minute of the actual goal $T_i$ in match $i$.\footnote{Note that this information becomes known only after the goal is scored.} To analyse potential shifts in betting odds and stakes placed in the minutes preceding the first goal, we consider only those 256 matches (covering 9,245 minutes with a 0:0 score) where the first goal was scored from minute 6 onwards.

Analysing potential anticipation of goals by market participants involves examining both bettors and bookmakers. Anticipating goals would result in shifts in odds set by bookmakers and changes in stakes placed by bettors, respectively. Bookmakers set their odds based on the expected outcome probability, often derived from forecasting models. Following previous literature (see, e.g.\ \citealp{feddersen2017sentiment,deutscher2018betting,winkelmann2021bookmakers}), we calculate implied winning probabilities from the betting odds, correcting for the bookmakers' margin, as follows:

$$\text{improb}_{itj} = \frac{1/O_{itj}}{1/O_{ith} + 1/O_{itd} + 1/O_{ita}},\quad j = h, d, a,$$

for a home win ($h$), a draw ($d$), and an away win ($a$). We denote the in-match implied probability for the team scoring the first goal as $\textit{improb}_{it}$. Typically, fixed match characteristics, such as information on the relative team strengths, potential home advantage, and possible injuries of players from both teams, are incorporated into the pre-match implied probabilities, denoted by $\textit{impprobpre}_{it}$. During the course of the match, betting markets potentially respond to in-match dynamics \citep{michels2023bettors}. Therefore, we consider the expected goals of both teams until minute $t$ of match $i$. Expected goals describe the number of goals to be expected given the scoring opportunities teams have had. As expected goals can be observed by bookmakers and bettors, they could potentially influence both betting odds and stakes placed. We denote the difference in (cumulative) expected goals between both teams from the beginning of the match until minute $t$ from the perspective of the team scoring the first goal by $\textit{xgdiff}_{it}$. We observe the difference in expected goals right before the first goal to range between -2.28 to 1.85, indicating that there are expected as well as surprising goals given the previous course of the match. On average, the expected goals for the scoring team are higher by 0.13 in the last minute before the goal occurs, suggesting that the team scoring the first goal had slightly more opportunities to score beforehand. Beyond goals, red cards constitute a very important event expected to affect (implied) winning probabilities. We capture red cards by the variables $\textit{redcardteam}_{it}$ (and $\textit{redcardopp}_{it}$), taking the value one if the team scoring the first goal (and the team conceding it, respectively) received a red card. Of the seven matches where a red card was issued before the first goal, the team with the numerical advantage scored the first goal in six. In our sample, we do not observe any matches where red cards were issued to both teams before the first goal was scored.

While the bookmaker's behaviour can be obtained from the odds offered, the stakes placed represent the bettors' behaviour. Since betting on draws is unpopular (in our dataset, only about 13\% of the money is placed on draws), we focus on the stakes placed on the team scoring the first goal relative to the total stakes placed on both teams at each minute, denoted by $\textit{stakerel}_{it}$. Analysing relative stakes (compared to other match outcomes) rather than absolute stakes avoids potential inaccuracies in the absolute stakes driven by closed markets, which can occur in case of penalty kicks or VAR decisions \citep{otting2024demand}. To investigate bettors' general behaviour, we examine descriptive statistics on the average relative stakes placed on the team eventually scoring the first goal throughout the scoreless period. Average relative stakes on the team scoring the first goal exhibit strong variation between 2.8\% and 97.1\%, with a mean of 58.1\% and a median of 65.0\%, indicating that bettors tend to place more money on teams eventually scoring the first goal. 

More specifically, we find that bettors prefer betting on the home team (53.4\% of relative stakes are placed on home teams and 46.6\% on away teams). This aligns with \citep{Staněk2017Home} and \citep{buhagiar2018some}, who report higher betting volumes on home teams. To account for bettors' behaviour, we introduce the variable $\text{\textit{home}}_t$ taking value 1 if we consider relative stakes placed on the home team in match $i$ (i.e.\ the home team scores the first goal in match $i$). Additionally, previous literature suggests that bettors tend to place more money on their favourite teams \citep{na2019not}. Consequently, it can be expected that the teams' sentiment influences the distribution of stakes between the two teams. As a proxy, we consider the variable $\text{\textit{volumediff}}_t$ denoting the difference in the average absolute stakes per minute placed on the two teams over all matches with a score of 0:0 contained in the dataset (see Table~\ref{tab:volume} in the Appendix for an overview of average stakes for each team).

\begin{table}[ht] 
\centering 
  \caption{Summary statistics of the key variables in the dataset.} 
  \label{tab:sum_stats} 
  \scalebox{0.8}{
\begin{tabular}{@{\extracolsep{5pt}} cccccc} 
\\[-1.8ex]\hline 
\hline \\[-1.8ex] 
Variable & Mean & Standard deviation & Minimum & Maximum & Median \\ 
\hline \\[-1.8ex] 
\textit{t} & 29.141 & 23.55 & 1 & 109 & 23 \\ 
\textit{mintogoal} & 29.141 & 23.55 & 1 & 109 & 23 \\ 
\textit{improb} & 0.435 & 0.175 & 0.046 & 0.922 & 0.416 \\ 
\textit{improbpre} & 0.467 & 0.188 & 0.048 & 0.915 & 0.432 \\ 
\textit{redcardteam} & 0.003 & 0.062 & 0 & 1 & 0 \\ 
\textit{redcardopp} & 0.025 & 0.155 & 0 & 1 & 0 \\ 
\textit{xgdiff} & 0.077 & 0.42 & -2.284 & 2.047 & 0.026 \\ 
\textit{home} & 0.573 & 0.495 & 0 & 1 & 1 \\ 
\textit{volumediff} & 6.881 & 16.425 & -46.715 & 46.773 & 7.583 \\ 
\textit{stakerel} & 0.581 & 0.288 & 0 & 1 & 0.651 \\ 
\hline \\[-1.8ex] 
\end{tabular}}
\end{table} 

Table~\ref{tab:sum_stats} presents summary statistics for the key variables in our dataset. We also examine the correlation coefficients between these variables (see Table~\ref{tab:correlation} in the Appendix). There are only a few correlations worth highlighting (absolute value of correlation coefficient larger than 0.35). Unsurprisingly, pre-match and in-match implied probabilities are highly correlated as long as the match remains scoreless. This is expected, as in the absence of major news, pre-match winning probabilities (a strong predictor for match outcomes) should align closely with in-match winning probabilities. Both variables are also correlated with the difference in the volume (\textit{volumediff}) and relative stakes (\textit{stakerel}), indicating that bettors tend to place more money on the favourite teams, according to implied probabilities. Furthermore, a positive difference in the average absolute stakes placed on a team over the whole season typically translates into higher relative stakes in a specific match. Except for the time variable, where we include linear, quadratic, and interaction terms, the variance inflation factors do not indicate multicollinearity among the covariates for the regression models considered in Section~\ref{sec:bookmakers}.

\begin{figure}[ht]
    \centering
    \includegraphics[width=0.9\linewidth]{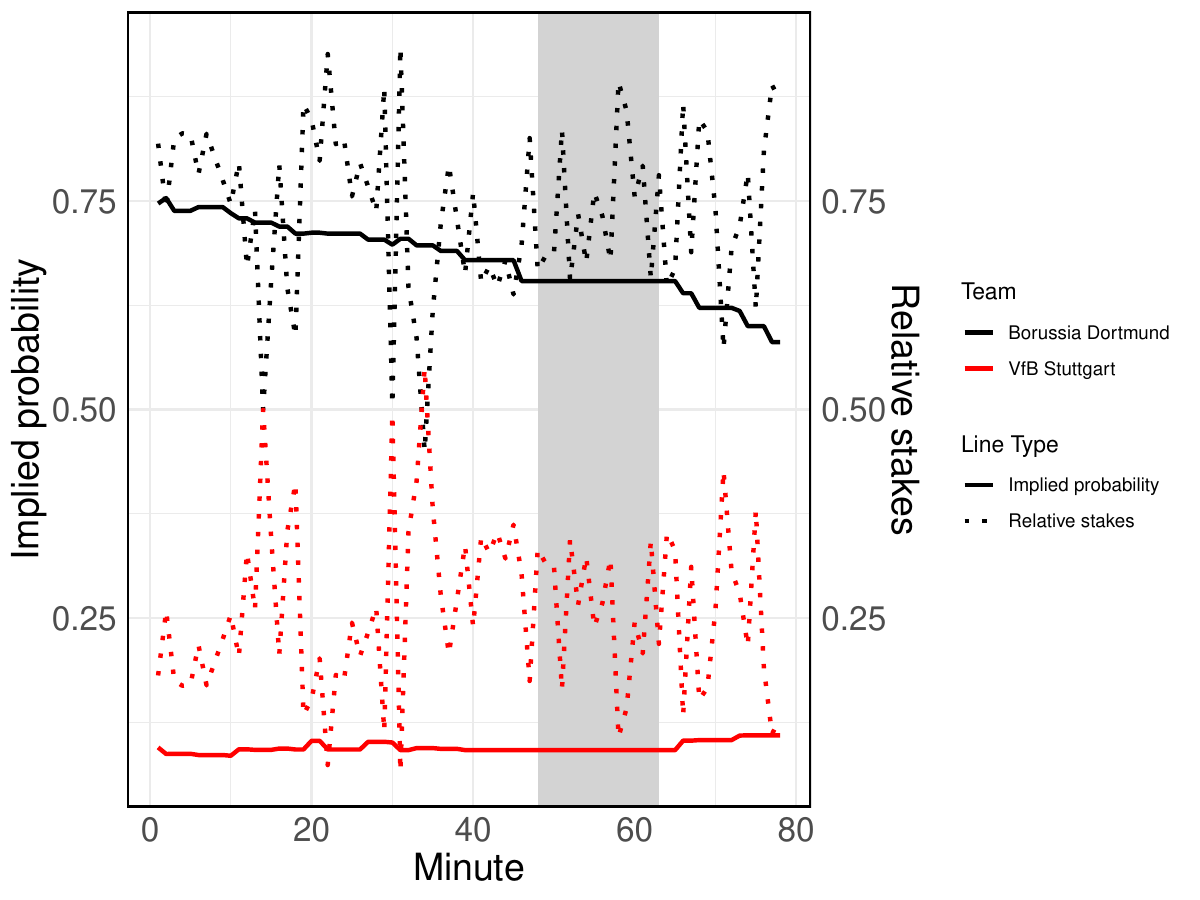}
    \caption{Implied probabilities (solid lines) and relative stakes (dotted lines) in the 2018/19 German Bundesliga match between Borussia Dortmund (black lines) and VfB Stuttgart (red lines) for the time period until the opening goal was scored by Dortmund. The gray hatched area corresponds to the half time break.}
    \label{fig:example_match}
\end{figure}

Figure~\ref{fig:example_match} illustrates relative stakes and implied probabilities for an example match between Borussia Dortmund and VfB Stuttgart in the 2018/19 Bundesliga season. The final result was 3:1, with Borussia Dortmund scoring the first goal 79 minutes after kick-off (in this case, minute 62 of the match). Borussia Dortmund was denoted to be the clear pre-match favourite with an implied winning probability (indicated by the red solid line) of 74.7\%. In contrast, VfB Stuttgart had a winning probability of only 9.5\% at the start of the match. While the implied probabilities for a Stuttgart win remained relatively constant during the scoreless period of the match, the implied probability for Dortmund decreased to 57.4\% before the first goal was scored. Although the relative stakes exhibited much more fluctuation throughout the match, stakes placed on Borussia Dortmund were, on average, about three times higher during the scoreless period of the match, with only one minute where higher stakes were placed on VfB Stuttgart.

\section{Do bookmakers anticipate goals?}
\label{sec:bookmakers}

This section examines whether bookmakers anticipate goals and adjust their odds accordingly. If bookmakers do anticipate goals, we would observe higher implied probabilities and, consequently, lower odds for the team that eventually scores the opening goal. Initially, we introduce basic model formulations that account for in-match dynamics. Finally, we explicitly analyse whether bookmakers anticipate goals and adjust the implied winning probabilities and odds.

\subsection{Basic model formulations}
To understand how bookmakers' odds evolve during a match, we formulate linear regression models to explain in-match implied win probabilities $\textit{improb}_{it}$ at minute $t$ for the team scoring the first goal in match $i$ by covariates. For the first model, we only consider the pre-match implied probabilities $\textit{improbpre}_{t}$ and propose the following formulation for the predictor of the first linear regression model (Model~\ref{eq:book1}):
\begin{equation}
    \begin{split}
        \nu_{it} = & \beta_0 + \beta_1 \cdot \text{\textit{improbpre}}_t
    \end{split}
    \label{eq:book1}
\end{equation}
with $E(\textit{improb}_{it}) = \nu_{it}$. Considering multiple observations for each match, we cluster standard errors at the match level to ensure correct p-values \citep{clse2, clse1}.

The left column of Table~\ref{tab:regmod123} presents the estimated coefficients and standard errors for the first model. We find a strong and statistically significant relationship between pre-match and in-match implied probabilities. However, on average, implied probabilities for the team scoring the first goal, before the goal occurs, are lower within the match than they are pre-match, as $\hat{\beta_0} + \hat{\beta_1} < 1$.

While pre-match probabilities comprehensively determine in-match implied probabilities at the start of a match, the probability of any team winning decreases (and the likelihood of a draw increases) as the match proceeds scoreless (see the example match in Figure~\ref{fig:example_match}). Consequently, we incorporate the minutes elapsed $t$ of match $i$ into the linear predictor. Given that this effect is expected to be stronger later in the match, we also include a quadratic effect of $t$. Furthermore, as observed in Figure~\ref{fig:example_match}, there is a stronger adjustment in implied probabilities during the match when pre-match implied probabilities are larger. Therefore, we include an interaction term between $t$ and $\textit{improbpre}_t$. This leads to the following formulation for Model~2:
\begin{equation}
    \begin{split}
        \nu_{it} = & \beta_0 + \beta_1 \cdot \text{\textit{improbpre}}_i + \beta_2 \cdot t_{it} + \beta_3 \cdot t_{it}^2 + \beta_4 \cdot \text{\textit{improbpre}}_i \cdot t_{it} \\
    \end{split}
    \label{eq:book2}
\end{equation}
In addition to the time elapsed, we expect major in-match events to influence the match outcome and, consequently, implied probabilities. Therefore, we include whether the team under consideration (and its opponent, respectively) has received a red card. Furthermore, we account for in-match dynamics by considering the difference in expected goals between the opponents, divided by the current minute $\text{\textit{xgdiff}}_{it}/t_{it}$. By dividing by the minute, we ensure that the value of the covariate can increase or decrease over time. Positive values correspond to situations where the team that eventually scores the first goal has had better goal-scoring opportunities. This leads to the formulation of Model~\ref{eq:book3}:
\begin{equation}
    \begin{split}
        \nu_{it} = & \beta_0 + \beta_1 \cdot \text{\textit{improbpre}}_i + \beta_2 \cdot t_{it} + \beta_3 \cdot t_{it}^2 + \beta_4 \cdot \text{\textit{improbpre}}_i \cdot t_{it} \\
        & + \beta_5 \cdot \text{\textit{redcardteam}}_{it} + \beta_6 \cdot \text{\textit{redcardopp}}_{it} + \beta_7 \cdot \text{\textit{xgdiff}}_{it}/t_{it}
    \end{split}
    \label{eq:book3}
\end{equation}
The second and third columns of Table~\ref{tab:regmod123} present results for the models that include in-match dynamics. All covariates have a statistically significant impact on the implied winning probability. Additionally, the Akaike Information Criterion (AIC) favours the most complex model (Model~3). In the initial minutes, in-match implied probabilities are almost entirely determined by the pre-match implied probabilities. However, as a scoreless match progresses, implied probabilities decrease, with the development depending on the pre-match implied probabilities, as indicated by the interaction term. This aligns with the prior expectation that draws become more likely the longer the score remains 0:0. Only for very small pre-match implied probabilities, we find in-match implied probabilities, on average, to increase very slightly for the first minutes of matches before they decrease again. Receiving a red card decreases implied winning probabilities by about 12 percentage points. In contrast, a red card of the opponent increases implied probabilities by approximately 17 percentage points. This difference occurs due to our perspective of the team that eventually takes the lead. If a team demonstrates higher expected goals per minute, this also increases the implied probability for this team, highlighting that bookmakers account for in-match dynamics.
\begin{table}[ht] \centering 
  \caption{Estimated coefficients and 95\%-confidence intervals for the basic linear regression model on bookmakers and models (Model~\ref{eq:book1}-~\ref{eq:book3}).} 
  \label{tab:regmod123} 
  \scalebox{0.6}{
\begin{tabular}{@{\extracolsep{5pt}}lccc} 
\\[-1.8ex]\hline 
\hline \\[-1.8ex] 
 & \multicolumn{3}{c}{\textit{Response variable:}} \\ 
\cline{2-4} 
\\[-1.8ex] & \multicolumn{3}{c}{Implied probability in-match} \\ 
\\[-1.8ex] & Model 1 & Model 2 & Model 3\\ 
\hline \\[-1.8ex] 
 \textit{Implied probability pre-match} & 0.885 & 1.016 & 1.003\\ 
  & [\,\,0.856, \,\,0.914] & [\,\,0.995, \,\,1.038] & [\,\,0.990, \,\,1.016] \\ 
  & \\ 
  \textit{Minute} & & 0.002 & 0.001 \\ 
  & & [\,\,0.001, \,\,0.002] & [\,\,0.001, \,\,0.002] \\ 
  & \\ 
  \textit{Minute$^2$} & & -0.000011 & -0.000013 \\ 
  & & [-0.000018, -0.000004] & [-0.000018, -0.000008] \\ 
  & \\ 
  \textit{Implied probability pre-match $\cdot$ Minute} & & -0.004 & -0.004 \\ 
  & & [-0.005, -0.003] & [-0.005, -0.003] \\ 
  & \\ 
  \textit{Red card team} & & & -0.120 \\ 
  & & & [-0.124, -0.116] \\ 
  & \\ 
  \textit{Red card opponent} & & & 0.173 \\ 
  & & & [\,\,0.136, \,\,0.210] \\ 
  & \\ 
  \textit{xgdiff per minute} & & & 0.163 \\ 
  & & & [\,\,0.075, \,\,0.250] \\ 
  & \\ 
  \textit{Constant} & \,\,0.022 & -0.010 & -0.004 \\ 
  & [\,\,0.009, \,\,0.035] & [-0.022, \,\,0.003] & [-0.011, \,\,0.004]\\ 
  & \\ 
\hline \\[-1.8ex] 
Observations & 9,425 & 9,425 & 9,425 \\ 
Akaike Inf. Crit. & -28,320 & -35,000 & -42,160 \\ 
\hline 
\hline \\[-1.8ex] 
\end{tabular}}
\end{table}

\subsection{Anticipation of goals}
\label{sec:book_anticipation}
The descriptive analysis indicates that the average difference in expected goals is slightly greater than zero just before the first goal occurs. This suggests that teams scoring the first goal have more opportunities to score before the initial goal. Simultaneously, bookmakers adjust their odds based on expected goals. This raises the question of whether bookmakers respond to in-match dynamics beyond the observable measure of expected goals. To investigate this, we extend Model~\ref{eq:book3} by including the covariate 1/\text{\textit{mintogoal}}$_{it}$. The value of this covariate increases before the first goal and reaches 1 in the minute before the goal occurs. Model~\ref{eq:book4} is formulated as follows:
\begin{equation}
    \begin{split}
        \nu_{it} = & \beta_0 + \beta_1 \cdot \text{\textit{improbpre}}_i + \beta_2 \cdot t_{it} + \beta_3 \cdot t_{it}^2 + \beta_4 \cdot \text{\textit{improbpre}}_i \cdot t_{it} \\
        & + \beta_5 \cdot \text{\textit{redcardteam}}_{it} + \beta_6 \cdot \text{\textit{redcardopp}}_{it} + \beta_7 \cdot \text{\textit{xgdiff}}_{it}/t_{it} \\
        & + \beta_8 \cdot \text{\textit{mintogoal}}^{-1}_{it} \\
    \end{split}
    \label{eq:book4}
\end{equation}
In the results (see Table~\ref{tab:regmod4}), we find that the parameter estimates are virtually unchanged from Model~\ref{eq:book3}. We observe a statistically insignificant effect for the remaining minutes until the goal is scored. This indicates that bookmakers do not anticipate goals beyond the observable expected goals and do not adjust their odds accordingly. This result remains robust even when the expected goals covariate is excluded. In summary, this section demonstrates that bookmakers adjust their implied outcome probabilities based on pre-match team strength and in-match information. However, they do not anticipate goals by adjusting probabilities before they occur. This potentially allows bettors to generate positive returns if they can anticipate goals. Whether bettors actually do anticipate goals is the subject of the next section.
\begin{table}[ht] \centering 
  \caption{Estimated coefficients and 95\%-confidence intervals for the linear regression models on bookmakers including remaining minutes to goal (Model~\ref{eq:book4}).} 
  \label{tab:regmod4} 
  \scalebox{0.6}{
\begin{tabular}{@{\extracolsep{5pt}}lc} 
\\[-1.8ex]\hline 
\hline \\[-1.8ex] 
 & \multicolumn{1}{c}{\textit{Response variable:}} \\ 
\cline{2-2} 
\\[-1.8ex] & \multicolumn{1}{c}{Implied probability in-match} \\ 
\\[-1.8ex] & Model 4 \\ 
\hline \\[-1.8ex] 
 \textit{Implied probability pre-match} & \,\,1.003\\ 
  & [\,\,0.990, 1.016] \\ 
  & \\ 
  \textit{Minute} & \,\,0.001 \\ 
  & [\,\,0.001, \,\,0.002] \\ 
  & \\ 
  \textit{Minute$^2$} & -0.000013 \\ 
  & [-0.000018, -0.000008] \\ 
  & \\ 
  \textit{Implied probability pre-match $\cdot$ Minute} & -0.004 \\ 
  & [-0.005, -0.003] \\ 
  & \\ 
  \textit{Red card team} & -0.120 \\ 
  & [-0.124, -0.116] \\ 
  & \\ 
  \textit{Red card opponent} & \,\,0.173 \\ 
  & [\,\,0.136, \,\,0.209] \\ 
  & \\ 
  \textit{xgdiff per minute} & \,\,0.165 \\ 
  & [\,\,0.077, \,\,0.253] \\ 
  & \\ 
  \textit{mintogoal}$^{-1}$ & -0.005 \\ 
  & [-0.012, \,\,0.002] \\ 
  & \\ 
  \textit{Constant} & -0.003 \\ 
  & [-0.010, \,\,0.004] \\ 
  & \\ 
\hline \\[-1.8ex] 
Observations & 9,425 \\ 
Akaike Inf. Crit. & -42,180 \\ 
\hline 
\hline \\[-1.8ex] 
\end{tabular}}
\end{table}

\section{Do bettors anticipate goals?}
\label{sec:bettors}

We now turn to the perspective of bettors placing stakes. To account for the highly dynamic betting market with serial correlation in the underlying activity level, we incorporate a latent variable capturing the current level of market activity. Initially, we aim to explain the relative stakes placed on the team scoring the first goal by the same covariates as those used for the bookmaker's implied probabilities. Finally, based on the suggestions of previous literature and our descriptive findings on bettors' behaviour, we additionally include the \textit{home} and \textit{volumediff} covariates to this model to enhance its explanatory power.

\subsection{Model formulation}
\label{sec:bettors_basic}
We consider the target variable $y_t$, representing the relative stakes placed on the team scoring the first goal in a match. As no bets can be placed when the market is closed (e.g.\ directly after a red card or after a penalty decision), we exclude those 60 observations from the analysis, reducing the total observations to 9,185 minutes. In the extreme case, all stakes placed during a given period are on one team. The target variable, hence, can take values between 0 and 1. Given this support of the variable, we apply a zero-one-inflated beta distribution \citep{ospina2012general,rigby2019distributions}:
\begin{align*}
    & y_t \sim \text{BEINF} (\mu_t, \sigma, \pi, \lambda) \\
    & f(y_t) = 
    \begin{cases}
        \pi & \text{if} \,\, y_t = 0\\
        (1 - \pi - \lambda) \cdot h(y_t) & \text{if} \,\, y_t \in (0,1)\\
        \lambda & \text{if} \,\, y_t = 1\\
    \end{cases}
\end{align*}
where $h(y_t)$ corresponds to the density function of a standard beta distribution. This model formulation aligns with previous work by \citet{michels2023bettors} for modelling relative stakes in betting markets. We parametrise the distribution in terms of its mean $\mu$ and a precision parameter $\gamma = \frac{\mu (1-\mu)}{\sigma^2} - 1$. We model the mean dependent on covariates and apply the inverse logit link function to the linear predictor: $\mu_{it} = \text{logit}^{-1}(\eta_{it})$.

Previous literature on finance and betting markets indicates the existence of an underlying (unobserved) market activity level that reflects the market's nervousness. Consequently, we include a latent state variable that captures market activity and model serial correlation in the state using an AR(1) process. Assuming a finite number of distinct states with clear interpretations, such as low and high levels of market activity, would lack the necessary flexibility to adequately model financial time series and capture gradual changes. Instead, we employ continuous-valued state-space models (SSMs) in discrete time, aligning with previous literature on stochastic volatility models for share returns \citep{langrock2012some,barra2017joint} and particularly stakes in betting markets \citep{michels2023bettors,otting2024demand}.

We consider the relative stakes placed on the team scoring the first goal in match $i$ at minute $t$, denoted $y_{it}$, as the state-dependent observation process $\{y_{it}\}$. This process is assumed to be driven by an underlying state process $\{s_{it}\}$. For simplicity, we will omit the match index $i$ in the following. The state process can be described as
$$s_t = \phi s_{t-1} + \sigma_s \epsilon_t$$
with $|\phi|$ as the persistence parameter, and $\sigma_s > 0$ and $\epsilon_t \sim N(0,1)$ specify the distribution of the error term. We assume the initial state $s_1$ to be generated by the stationary distribution of the state process: $\delta \sim N \left(0, \sqrt{\frac{\sigma_s^2}{1-\phi^2}}\right)$. The state active at observation $t$ determines the current state-dependent distribution. The linear predictor for the mean of the observation process in the SSM $\eta_t$ is extended by the state variable: $\eta_t = \nu_t + s_t$. The dependence structure of the SSM is illustrated in Figure~\ref{fig:SSM} (see Appendix~\ref{sec:SSM_implementation} for technical details on the implementation).

\begin{figure}[!ht]
    \centering
    \scalebox{0.65}{
	\begin{tikzpicture}
	\node[circle,draw=black, fill=gray!5, inner sep=0pt, minimum size=50pt] (A) at (-2, -3) {...};
	\node[circle,draw=black, fill=gray!5, inner sep=0pt, minimum size=50pt] (B) at (2, -3) {$s_{t-1}$};
	\node[circle,draw=black, fill=gray!5, inner sep=0pt, minimum size=50pt] (C) at (6, -3) {$s_{t}$};
	\node[circle,draw=black, fill=gray!5, inner sep=0pt, minimum size=50pt] (D) at (10, -3) {$s_{t+1}$};
 	\node[circle,draw=black, fill=gray!5, inner sep=0pt, minimum size=50pt] (E) at (14, -3) {...};
   	\node[circle,draw=black, fill=gray!5, inner sep=0pt, minimum size=50pt] (F) at (2, 0) {$y_{t-1}$};
	\node[circle,draw=black, fill=gray!5, inner sep=0pt, minimum size=50pt] (G) at (6, 0) {$y_{t}$};
	\node[circle,draw=black, fill=gray!5, inner sep=0pt, minimum size=50pt] (H) at (10, 0) {$y_{t+1}$};
   	\node[circle,draw=black, fill=gray!5, inner sep=0pt, minimum size=50pt] (I) at (2, 3) {$x_{t-1}$};
	\node[circle,draw=black, fill=gray!5, inner sep=0pt, minimum size=50pt] (J) at (6, 3) {$x_{t}$};
	\node[circle,draw=black, fill=gray!5, inner sep=0pt, minimum size=50pt] (K) at (10, 3) {$x_{t+1}$};
    \draw[-{Latex[scale=2]}] (A)--(B);
	\draw[-{Latex[scale=2]}] (B)--(C);
    \draw[-{Latex[scale=2]}] (C)--(D);
    \draw[-{Latex[scale=2]}] (D)--(E);
    \draw[-{Latex[scale=2]}] (B)--(F);
    \draw[-{Latex[scale=2]}] (C)--(G);
    \draw[-{Latex[scale=2]}] (D)--(H);
    \draw[-{Latex[scale=2]}] (I)--(F);
    \draw[-{Latex[scale=2]}] (J)--(G);
    \draw[-{Latex[scale=2]}] (K)--(H);
	\end{tikzpicture}}
\caption{Dependence structure of the SSM with latent state $s_t$, covariates $x_t$ and response variable $y_t$.}
\label{fig:SSM}
\end{figure}
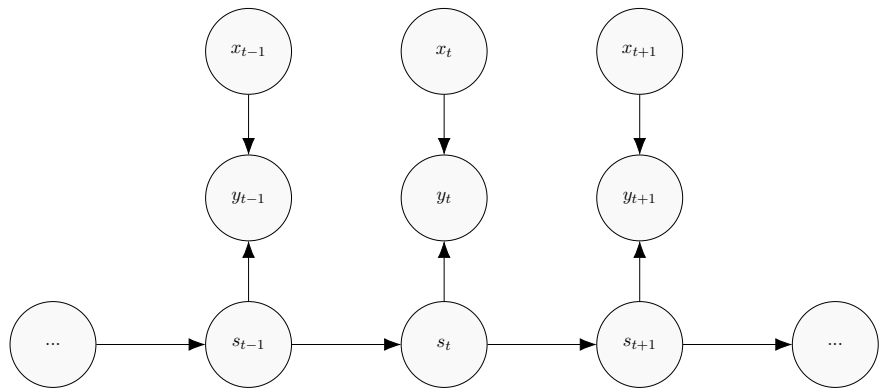 

\subsection{Results}
Table~\ref{tab:bettors_extended} presents the estimated coefficients along with their confidence intervals. Firstly, we consider the same covariates as in the final bookmaker's model in Section~\ref{sec:book_anticipation}, but extended by the state process. Parameter estimates for this SSM are provided in the left column of Table~\ref{tab:bettors_extended}. Results for the state process, with $\hat{\phi}$ close to 1, indicate strong serial correlation in the underlying market activity level. Indeed, the AIC clearly favours this model over a model without state process ($\Delta \text{AIC} = 9,355$; see Appendix~\ref{sec:mod_bettor_without} for a model without state process). 

Estimates for the state-dependent covariates confirm descriptive findings that bettors tend to place higher (relative) stakes on favourites with higher (pre-match) implied winning probabilities, as implied by the bookmaker’s odds. Simultaneously, the results suggest that the distribution of relative stakes between both teams does not depend on the minute (neither in the linear form nor the quadratic and interaction terms). While the effect of a red card for the team scoring the goal is negative but insignificant, relative stakes significantly increase when the opponent receives a red card. This also applies to teams showing a comparative advantage during the match, as indicated by the difference in expected goals per minute. Regarding the potential anticipation of goals by bettors, the model does not provide evidence for adjusted behaviour of bettors before the first goal. The estimated coefficients $\hat{\pi} = 0.00096$ and $\hat{\lambda} = 0.00053$ are close to the empirical probabilities that all stakes are placed on the team (and the opponent, respectively) in a given minute. The precision parameter is estimated as $\hat{\gamma} = 16.065$.

Following the descriptive analysis on bettors' tendency to place higher stakes on home teams and those with higher sentiment, we additionally include the corresponding covariates in the final model. Simultaneously, the time-dependent covariates appear insignificant in the basic model. Therefore, we exclude the additional covariates (quadratic effect and interaction term) from the final model formulation (for the full model including all covariates, see Table~\ref{tab:bettors_final} in the Appendix). The linear predictor for the final model is given as follows:
\begin{equation}
    \begin{split}
        \nu_{it} = & \beta_0 + \beta_1 \cdot \text{\textit{improbpre}}_i + \beta_2 \cdot t_{it} + \beta_3 \cdot \text{\textit{redcardteam}}_{it} + \beta_4 \cdot \text{\textit{redcardopp}}_{it} \\ 
        & + \beta_5 \cdot \text{\textit{home}}_{t} + \beta_6 \cdot \text{\textit{volumediff}}_{t} + \beta_7 \cdot \text{\textit{xgdiff}}_{it}/t_{it} + \beta_8 \cdot \text{\textit{mintogoal}}^{-1}_{it}
    \end{split}
\end{equation}

\begin{table}[ht!] \centering 
      \caption{Estimated coefficients and 95\%-confidence intervals for the state-space models (SSMs) with beta distribution for bettors.} 
    \label{tab:bettors_extended}
    \scalebox{0.6}{
    \begin{tabular}{@{\extracolsep{5pt}}lcc} 
        \\[-1.8ex]\hline 
        \hline \\[-1.8ex] 
         & \multicolumn{2}{c}{\textit{Response variable:}} \\ 
        \cline{2-3} 
        \\[-1.8ex] & \multicolumn{2}{c}{Relative stakes team} \\ 
        \\[-1.8ex] & Basic SSM & Final SSM \\ 
        \hline \\[-1.8ex] 
         $\phi$ & \,\,0.984 & \,\,0.974 \\
         & [\,\,0.981 \,\,0.987] & [\,\,0.972, \,\,0.977]\\
         $\sigma_s$ & \,\,0.176 & \,\,0.183 \\ 
         & [\,\,0.165 \,\,0.188] & [\,\,0.176, \,\,0.190]\\
        \hline \\[-1.8ex] 
         \textit{Implied probability pre-match} & 4.520 & 1.753 \\ 
          & [\,\,3.990, \,\,5.051] & [\,\,1.648, \,\,1.859] \\ 
          & \\ 
          \textit{Minute} & \,\,0.002 & \,\,0.000 \\ 
          & [-0.007, \,\,0.011] & [-0.002, \,\,0.003] \\ 
          & \\ 
          \textit{Minute$^2$} & 0.00001 &  \\ 
          & [-0.00006, 0.00009] &  \\ 
          & \\ 
          \textit{Implied probability pre-match $\cdot$ Minute} & -0.004 &  \\ 
          & [-0.020, \,\,0.013] &  \\ 
          & \\ 
          \textit{Red card team} & -0.630 & -0.767 \\ 
          & [-1.499, \,\,0.240] & [-1.346, -0.188] \\ 
          & \\ 
          \textit{Red card opponent} & \,\,0.661 & \,\,0.681 \\ 
          & [\,\,0.389, \,\,0.932] & [\,\,0.215, \,\,1.147] \\ 
          & \\ 
          \textit{xgdiff per minute} & \,\,3.497 & \,\,3.627 \\ 
          & [\,\,3.118, \,\,3.875] & [\,\,3.552, \,\,3.702] \\ 
          & \\ 
          \textit{home} &  & -0.005 \\
          &  & [-0.185, \,\,0.175]\\
          & \\ 
          \textit{volumediff} &  & \,\,0.047 \\
          &  & [\,\,0.042, \,\,0.051]\\
          & \\ 
          \textit{mintogoal}$^{-1}$ & \,\,0.089 & \,\,0.096 \\
          & [-0.159, \,\,0.336] & [-0.031, \,\,0.222]\\
          & \\ 
          \textit{Constant} & -1.785 & -0.762 \\
          & [-2.058, -1.513] & [-0.969, -0.555]\\
          & \\ 
        \hline \\[-1.8ex] 
        Observations & 9,185 & 9,185\\ 
        Akaike Inf. Crit. & -14,581.68 & -14,715.64 \\ 
        \hline 
        \hline \\[-1.8ex] 
    \end{tabular}}
\end{table}

Results in the right column of Table~\ref{tab:bettors_extended} confirm the high serial correlation in the market activity level. In this final model, the effect of red cards for both teams is significant. The findings corroborate bettors' tendency to place higher relative stakes on teams with higher sentiment. However, higher relative stakes on the home teams are not verified, given that we control for pre-match implied probabilities and the difference in the average betting volume between both teams. Again, there is no evidence for anticipation observed in bettors' behaviour. While applying various models throughout this section, the findings on the potential anticipation of goals are robust: relative stakes placed on the team eventually scoring the goal do not increase prior to its occurrence.

\section{Discussion}
\label{sec:conclusion}

In this paper, we explore whether market participants anticipate the first goal of a match by adjusting odds and stakes toward the team eventually scoring the goal. Although major events, such as red cards and in-match dynamics reflected by expected goals, significantly impact the behaviour of both sides of the market, our findings reveal that neither odds nor stakes show significant adjustments right before the first goal. Thus, we can conclude that neither bookmakers nor bettors anticipate the first goal of a match.

The results add important knowledge for financial markets and betting markets in particular. They demonstrate that neither professional bookmakers nor (predominantly non-professional) bettors can anticipate the occurrence of major news. In comparison to financial markets, football betting benefits from the absence of insider trading. Apart from match-fixing \citep{otting2018integrating}, which was not detected in the season considered in this paper, both market sides adjust their actions in response to processing the information available. Hence, our findings can clearly differentiate between insider trading and market anticipation, a distinction that is challenging in financial markets \citep{jain2014stock, tunyi2021revisiting}. Bookmaker odds are rigid as we do not observe adjustments beyond expected goals and red cards. Conversely, (relative) stakes placed by bettors are much more volatile. Yet, we identify an underlying latent variable capturing the level of market activity to exhibit strong serial correlation. Arbitrage would occur if bettors were able to predict goals; however, our results indicate that bettors do not anticipate goals by increasing stakes on the teams scoring the first goal. This suggests that bookmakers do not face negative consequences from failing to anticipate goals.

While the setting of the German Bundesliga appears advantageous, there is also a drawback. The high level of information availability could potentially reduce market inefficiencies. \citet{elaad2020information} show that bookmaker margins are smaller for higher leagues. This is potentially driven by stronger competition among bookmakers in higher tiers or increased market transparency due to greater media coverage. Following the argument that less information is available for smaller leagues, implicit knowledge of bettors could be more valuable, possibly leading to a stronger anticipation effect of goals by fans. Additionally, our findings form the basis for further investigations into in-match betting. The risk of match-fixing has been a subject of interest in the literature for several years \citep{otting2018integrating,forrest2019using}. As our results indicate no significant adjustments towards the team eventually scoring the first goal, observing such anticipatory behaviour and a significant change in stakes placed on the team could potentially signal fraudulent activity. However, to analyse whether our model could correctly identify fixed matches, we would need to extend our analysis to matches where fixing was proven. Furthermore, absolute stakes should also be considered in such an analysis.

\newpage
\bibliographystyle{apalike} 
\bibliography{library}

\newpage
\section*{Appendix}

\begin{appendices}

\section{Average absolute stakes per minute placed on teams}
\begin{table}[htp!]
    \caption{Average absolute stakes per minute placed on the 18 German Bundesliga teams in the 2018/19 season for all matches during the time period with a score of 0:0. Note that all actual stakes are transformed by the same fixed constant, since we are not allowed to provide any information on the actual money placed.}
    \label{tab:volume}
    \centering
    \begin{tabular}{l|l}
        Team & Average stakes per minute \\
        \hline
        Borussia Dortmund & 55.70 \\
        Borussia M'gladbach & 36.80 \\
        Eintracht Frankfurt & 34.68 \\
        Bayern München & 32.99 \\
        RB Leipzig & 27.55 \\
        Schalke 04 & 25.51 \\
        Bayer Leverkusen & 25.50 \\
        1899 Hoffenheim & 23.20 \\
        Werder Bremen & 20.88 \\
        Hertha BSC & 17.26 \\
        VfL Wolfsburg & 16.51 \\
        VfB Stuttgart & 12.57 \\
        FSV Mainz 05 & 11.23 \\
        FC Augsburg & 10.25 \\
        SC Freiburg & \,\,\,9.84 \\
        Hannover 96 & \,\,\,9.05 \\
        Fortuna Düsseldorf & \,\,\,8.98 \\
        1. FC Nürnberg & \,\,\,8.93 \\
        \end{tabular}
\end{table}

\newpage
\section{Correlation coefficients between key variables}

\begin{table}[ht]
\centering
\caption{Correlation coefficients between the key variables in the dataset.} 
\label{tab:correlation}
\begin{sideways}
\scalebox{0.58}{
\begin{tabular}{@{\extracolsep{5pt}} rrrrrrrrrrr} 
\\[-1.8ex]\hline 
\hline \\[-1.8ex] 
 & \textit{t} & \textit{mintogoal} & \textit{improb} & \textit{improbpre} & \textit{redcardteam} & \textit{redcardopp} & \textit{xgdiff} & \textit{home} & \textit{volumediff} & \textit{stakerel} 
 \\ 
\hline \\[-1.8ex] 
\textit{nrinterval} & 1 & $-$0.239 & $-$0.157 & 0.031 & $-$0.017 & 0.168 & 0.214 & 0.046 & 0.056 & 0.058 \\ 
\textit{mintogoal} & $-$0.239 & 1 & 0.086 & 0.031 & $-$0.028 & $-$0.036 & $-$0.077 & 0.046 & 0.056 & 0.028 \\ 
\textit{improb} & $-$0.157 & 0.086 & 1 & 0.951 & $-$0.119 & 0.125 & 0.270 & 0.325 & 0.656 & 0.599 \\ 
\textit{improbpre} & 0.031 & 0.031 & 0.951 & 1 & $-$0.081 & 0.006 & 0.307 & 0.334 & 0.704 & 0.619 \\ 
\textit{redcardteam} & $-$0.017 & $-$0.028 & $-$0.119 & $-$0.081 & 1 & $-$0.010 & $-$0.042 & $-$0.072 & $-$0.026 & $-$0.089 \\ 
\textit{redcardopp} & 0.168 & $-$0.036 & 0.125 & 0.006 & $-$0.010 & 1 & 0.048 & 0.101 & $-$0.029 & 0.054 \\ 
\textit{xgdiff} & 0.214 & $-$0.077 & 0.270 & 0.307 & $-$0.042 & 0.048 & 1 & 0.120 & 0.265 & 0.328 \\ 
\textit{home} & 0.046 & 0.046 & 0.325 & 0.334 & $-$0.072 & 0.101 & 0.120 & 1 & 0.011 & 0.081 \\ 
\textit{volumediff} & 0.056 & 0.056 & 0.656 & 0.704 & $-$0.026 & $-$0.029 & 0.265 & 0.011 & 1 & 0.686 \\ 
\textit{stakerel} & 0.058 & 0.028 & 0.599 & 0.619 & $-$0.089 & 0.054 & 0.328 & 0.081 & 0.686 & 1 \\ 
\hline \\[-1.8ex] 
\end{tabular}}
\end{sideways}
\end{table}

\newpage

\section{Technical details on the SSM implementation}
\label{sec:SSM_implementation}

We estimate model parameters using the maximum likelihood method. The analytical evaluation of this likelihood is typically intractable, as it involves $T$-dimensional integrals, where $T$ corresponds to the number of observations. Instead, we rely on approximation methods as introduced by \citet{kitagawa1987non}. The exact likelihood, given the model structure incorporating the Markov property of the state process and conditional independence of observations (see Figure~\ref{fig:SSM}), can be written as follows: 
\begin{equation}
\mathcal{L}_T (\bm{\Theta}) = \int\cdots\int f(s_1)f(y_1|s_1)\prod\limits_{t=2}^{T} f(s_t|s_{t-1})f(y_t|s_t)ds_T\ldots ds_1.\\
\end{equation}
\label{SSMllk}    
Here, $\Theta$ denotes the parameter vector. We discretise the state space by defining $m$ intervals $B_i=(b_{i-1},b_i)$ with length $h=(b_m-b_0)/m$ and midpoints $b_i^*=\frac{b_i-b_{i-1}}{2}$. If $m$ is chosen sufficiently large, we can approximate the likelihood up to some decimal places as follows \citep{langrock2012some}:
\begin{equation}
{\mathcal{L}}_T (\bm{\Theta}) \approx h^T\sum\limits_{i_1=1}^m\cdots\sum\limits_{i_{T}=1}^m f(b_{i_1}^*)f(y_1|b_{i_1}^*)\prod\limits_{t=2}^{T}f(b_{i_t}^*|b_{i_{t-1}}^*)f(y_t|b_{i_t}^*).
\label{eq:llk}
\end{equation}
This approximated likelihood can be calculated at computational costs of order $\mathcal{O}(Tm^2)$ when using the forward algorithm (\citealp{ZML}, Chapter 11). In fact, the likelihood reduces to that of an $m$-state hidden Markov model (HMM), enabling the application of well-known HMM tools. The transition probability matrix $\Gamma$, encompassing the probabilities of switching from one state to another $\gamma_{ij} = f(s_t = b^*_j | s_{t-1} = b_i^*), i,j=1,\ldots, m$ is fully determined by $\phi$ and $\sigma_s$. The state-dependent densities $f(y_t | s_t = b_i)$ are provided by the $m \times m$ diagonal matrix $\mathbf{P}(y_t)$. We can now reformulate and recursively calculate the (approximate) likelihood stated in Equation~(\ref{eq:llk}) by
\begin{equation*}
{\mathcal{L}}_T (\bm{\Theta}) \approx \bm{\delta} \mathbf{P}(y_1) \bm{\Gamma} \mathbf{P}(y_2) \ldots \bm{\Gamma} \mathbf{P}(y_{T}) \bm{1}
\end{equation*}
\noindent
with $\bm{1} = (1,\ldots,1)' \in \mathbb{R}^m$. 

\newpage
For the given longitudinal dataset, we assume independence between matches and calculate the total likelihood as the product of individual likelihoods for each match. We employ the Broyden–Fletcher–Goldfarb–Shanno (BFGS) algorithm, a quasi-Newton method for numerical maximisation, to obtain parameter estimates in Python, subject to technical details \citep{ZML}. To balance the trade-off between computation time and an accurate discretisation of the state space, we use $m = 95$ intervals, which is close to the conservative choice proposed in, for example, \citet{langrock2013maximum,mews2024maximum}, with bounds $-b_0 = b_m = 3$.

\section{Bettors' model without state process}
\label{sec:mod_bettor_without}

\begin{table}[!htbp] \centering 
  \caption{Estimated coefficients and 95\%-confidence intervals for the beta regression model on bettors without state process.} 
  \label{tab:bettors1}
  \scalebox{0.6}{
\begin{tabular}{@{\extracolsep{5pt}}lc} 
\\[-1.8ex]\hline 
\hline \\[-1.8ex] 
 & \multicolumn{1}{c}{\textit{Response variable:}} \\ 
\cline{2-2} 
\\[-1.8ex] & \multicolumn{1}{c}{Relative stakes team} \\ 
\hline \\[-1.8ex] 
 \textit{Implied probability pre-match} & 4.333 \\ 
  & [\,\,4.148, \,\,4.517] \\ 
  & \\ 
  \textit{Minute} & \,\,0.008 \\ 
  & [\,\,0.005, \,\,0.011] \\ 
  & \\ 
  \textit{Minute$^2$} & -0.00001 \\ 
  & [-0.00004, 0.00003] \\ 
  & \\ 
  \textit{Implied probability pre-match $\cdot$ Minute} & -0.014 \\ 
  & [-0.018, -0.009] \\ 
  & \\ 
  \textit{Red card team} & -0.591 \\ 
  & [-0.904, -0.277] \\ 
  & \\ 
  \textit{Red card opponent} & \,\,0.418 \\ 
  & [\,\,0.291, \,\,0.545] \\ 
  & \\ 
  \textit{xgdiff per minute} & \,\,7.961 \\ 
  & [\,\,7.508, \,\,8.414] \\ 
  & \\ 
  \textit{mintogoal}$^{-1}$ & -0.109 \\ 
  & [-0.216, -0.003] \\ 
  & \\ 
  \textit{Constant} & -1.700 \\ 
  & [-1.788, -1.611] \\ 
  & \\ 
\hline \\[-1.8ex] 
Observations & 9,185 \\ 
Akaike Inf. Crit. & -5,226.72 \\ 
\hline 
\hline \\[-1.8ex] 
\end{tabular}}
\end{table}

\newpage

\section{Results for the full bettors' model}

\begin{table}[ht] \centering 
      \caption{Estimated coefficients and 95\%-confidence intervals for the state-space model (SSM) with beta distribution for bettors including all variables.} 
    \label{tab:bettors_final}
    \scalebox{0.6}{
    \begin{tabular}{@{\extracolsep{5pt}}lc} 
        \\[-1.8ex]\hline 
        \hline \\[-1.8ex] 
         & \multicolumn{1}{c}{\textit{Response variable:}} \\ 
        \cline{2-2} 
        \\[-1.8ex] & \multicolumn{1}{c}{Relative stakes team} \\ 
        \hline \\[-1.8ex] 
          $\phi$ & \,\,0.974 \\
         & [\,\,0.970, \,\,0.978]\\
         & \\
        $\sigma_s$ & \,\,0.183 \\
        & [\,\,0.174, \,\,0.193]\\
        \hline \\[-1.8ex] 
         \textit{Implied probability pre-match} & \,\,1.929 \\
         & [\,\,1.594, \,\,2.263]\\
        & \\
        \textit{Minute} & \,\,0.004 \\
        & [-0.004, \,\,0.012]\\
        & \\
        \textit{Minute$^2$} & \,\,0.00000 \\
        & [-0.00007, 0.00008]\\
        & \\
        \textit{Implied probability pre$-$match $\cdot$ Minute} & -0.007 \\
        & [-0.022, \,\,0.008]\\
        & \\
        \textit{Red card team} & -0.777 \\
        & [-1.601, \,\,0.047]\\
        & \\
        \textit{Red card opponent} & \,\,0.682 \\
        & [\,\,0.310, \,\,1.054]\\
        & \\
        \textit{xgdiff per minute} & \,\,3.640 \\
        & [\,\,1.641, \,\,5.639]\\
        & \\
        \textit{Home} & -0.009 \\
        & [-0.178, \,\,0.161]\\
        & \\
        \textit{Volume} & \,\,0.047 \\
        & [\,\,0.041, \,\,0.052]\\
        & \\
        \textit{mintogoal}$^{-1}$ & \,\,0.095 \\
        & [-0.454, \,\,0.645]\\
        & \\
        \textit{Constant} & -0.842 \\
        & [-1.052, -0.631]\\
        & \\
        \hline \\[-1.8ex] 
        Observations & 9,185 \\ 
        Akaike Inf. Crit. & -14,712.72\\ 
        \hline 
        \hline \\[-1.8ex] 
    \end{tabular}}
\end{table}

\end{appendices}

\end{spacing}
\end{document}